\documentclass[preprintnumbers, aps, floatfix, onecolumn, preprintnumbers, letterpaper, superscriptaddress,nofootinbib,10pt]{revtex4-2}
\usepackage{dcolumn}
\usepackage{graphicx}
\usepackage{amsmath}
\usepackage{amsfonts}
\usepackage{amssymb}
\usepackage{microtype}
\usepackage{subfigure}
\usepackage{makeidx}
\usepackage{bm}
\usepackage{epsf}
\usepackage[colorlinks=true,citecolor=blue,linkcolor=black,urlcolor=black]{hyperref}
\usepackage{color}
\usepackage{multirow,dcolumn}
\usepackage{graphicx}
\usepackage{mathrsfs}
\graphicspath{{Images/}}

\def\doi{http://doi.org}

\def\be{\begin{equation*}}
\def\ee{\end{equation*}}

\def\Ref{\ref}

\begin{document}

\title{Rotating $(2+1)$-dimensional Black Holes in Einstein-Maxwell-Dilaton Theory}
\author{Thanasis Karakasis}
\email{thanasiskarakasis@mail.ntua.gr} \affiliation{Physics Division,
National Technical University of Athens, 15780 Zografou Campus,
Athens, Greece.}

\author{Eleftherios Papantonopoulos}
\email{lpapa@central.ntua.gr} \affiliation{Physics Division,
National Technical University of Athens, 15780 Zografou Campus,
Athens, Greece.}

\author{Zi-Yu Tang}
\email{tangziyu@ucas.ac.cn}
\affiliation{School of Fundamental Physics and Mathematical Sciences, Hangzhou Institute for Advanced Study, UCAS, Hangzhou 310024, China}
\affiliation{School of Physical Sciences, University of Chinese Academy of Sciences, Beijing 100049, China}

\author{Bin Wang}
\email{wang\_b@sjtu.edu.cn}
\affiliation{Center for Gravitation and Cosmology, College of Physical Science
and Technology, Yangzhou University, Yangzhou 225009, China}
\affiliation{School of Aeronautics and Astronautics, Shanghai Jiao Tong University, Shanghai 200240, China}

\begin{abstract}

We consider Einstein-Maxwell-Dilaton theory in $(2+1)$-dimensions where the coupling between the scalar field and the Maxwell invariant is the dilatonic coupling $f(\phi) = \exp (-2\alpha \phi)$  and obtain novel exact rotating black hole solutions. The dilatonic parameter $\alpha$ impacts the metric function, affecting  the rotating properties of the black hole, its mass and also its thermodynamics. Calculating the  entropy we find that it is always positive and the dilatonic black holes may have higher entropy than the BTZ black hole. Depending on the parameters, the dilatonic BTZ-like black hole may be thermodynamically preferred than the BTZ black hole which is recovered when $\alpha=0$.
\end{abstract}

\maketitle

\tableofcontents

\section{Introduction}
 In $(2+1)$-dimensions one of the first exact black holes with a negative cosmological constant was discussed by Ba\~{n}ados,  Teitelboim, and Zanelli (BTZ) \cite{BTZ1,BTZ2}. This spacetime is a solution of a theory that consists of the Ricci scalar, a negative cosmological constant and the electromagnetic invariant which can be linear \cite{BTZ1,BTZ2,Martinez:1999qi} or non-linearly coupled to gravity \cite{nonlinear} which are regular solutions having a non-singular center. Regular solutions were found recently in \cite{Bueno:2021krl} involving a scalar field.  In \cite{Martinez:1996gn,Henneaux:2002wm} $(2+1)$-dimensional black holes with a conformally coupled scalar field, being regular everywhere, were discussed.  After these first results other hairy black holes in $(2+1)${\color{red}-}dimensions were discussed  \cite{CMT1,CMT2,Correa:2012rc,Natsuume,Aparicio:2012yq,Xu:2014uha}. In \cite{Cardenas:2014kaa}  $(2+1)$-dimensional gravity with negative cosmological constant in the presence of a scalar field and an Abelian gauge field was considered. Both fields are conformally coupled to gravity, the scalar field through a non-minimal coupling to the curvature and the gauge field given by a power of the Maxwell field. A sixth-power self-interacting potential, which does not spoil conformal invariance is also included in the action, introducing a relation between the scalar curvature and the cosmological constant. In \cite{Xu:2014uka} and \cite{Xu:2013nia} $(2+1)$-dimensional charged black holes with scalar hairs were derived, where the scalar potential is not fixed ad hoc but  derived from Einstein's equations. In \cite{Tang:2019jkn} exact three dimensional black holes with non-minimal coupling scalar field were discussed, while $(2+1)$-dimensional $f(R)$ gravity black holes coupled to scalar fields were also obtained \cite{Karakasis:2021lnq,Karakasis:2021ttn}.

In $(3+1)$-dimensions there is also a rich literature in black holes coupled to scalar fields. The first attempts were carried out by Bekenstein, Bocharova, Bronnikov and  Melnikov called BBMB black hole \cite{BBMB}. The resulting spacetime resembles the extremal Reissner-Nordstr\"om one but the scalar field diverges at the horizon and it was also shown in \cite{bronnikov} that this solution is unstable under scalar perturbations. It was shown that the resultant temperature is zero and the entropy diverges allowing   to argue that the BBMB solution does not describe a black hole spacetime, or that the thermodynamics of this solution is not well understood \cite{Zaslavskii:2002zv}. To make the scalar field regular at the horizon a gravitational scale was introduced via a cosmological constant \cite{Martinez:2002ru}, resulting to an asymptotically de Sitter spacetime which is also found to be unstable \cite{Harper:2003wt}. The thermodynamics of this solution was also found to have an odd behavior \cite{Barlow:2005yd, Winstanley:2004ay}. Recently, another gravitational scale was introduced via the modification of the curvature part of the action by adding a non-linear function of the Ricci scalar ($f(R)$ gravity) \cite{Karakasis:2021rpn} making the scalar field regular at the horizon. The thermodynamics of both the general relativity and $f(R)$ gravity cases also suffer from the same problems (negative entropy) and in both cases an electric charge has to be introduced to cure this problem \cite{Barlow:2005yd, Winstanley:2004ay,Karakasis:2021rpn}. Black holes with scalar fields, minimally or non-minimally coupled to gravity and other matter fields have been extensively  discussed in the literature \cite{Dotti:2007cp}-\cite{Rinaldi:2012vy}.

A dilatonic field appears naturally in the low energy limit of various string theories \cite{stringdilaton}, in $(3+1)$-dimensions. As a result black hole solutions were obtained in these theories \cite{blackstring}. These pioneering results, showed that the produced compact object is determined by the black hole mass, the electric charge and the asymptotic value of the dilatonic scalar field. The scalar field dresses the black hole with a secondary scalar hair and there is no independent scalar charge that can be detected asymptotically by an observer, since the scalar charge is given in terms of the conserved quantities of the black hole, i.e. the mass and the electric charge. Black holes in $(3+1)$-dimensions with a dilatonic coupling has been a very active field of research, and different scenarios with self-interacting scalar potantials, different types of electrodynamics and Yang-Mills fields, and spacetimes with different types of topology have been studied through the years \cite{Myers:1987qx}-\cite{Priyadarshinee:2021rch}.

Dilatonic black hole solutions were also found in $(2+1)$-dimensions. Modification of the BTZ black hole by a dilaton scalar field has been considered in \cite{Chan:1996rd}, where it was found that a second degree of freedom has to be added in the metric ansatz to obtain  non-trivial solutions which is expected, since a simple self-interacting scalar field theory does not satisfy $\rho=-p_r$ \cite{Jacobson:2007tj}. This work was further extended  in \cite{Chan:1994qa} in which a dilaton field $\phi=ln(r\beta)$ coupled to an electromagnetic field was introduced and in the presence of a potential $V(\phi)=\exp(b\phi)\Lambda$ the causal structures were investigated, and  the thermodynamical temperature and entropy were computed. In \cite{Chan:1995wj} using the same forms of the dilaton field and the potential term spinning black hole solutions were obtained, where the coupling term does not account since the solutions are uncharged.

Black holes were also investigated in \cite{Dehghani:2017thu} where a self-interacting potential that is a linear combination of two Liouville type potentials was considered. One can also consider different types of electrodynamics to find black hole solutions as in \cite{Yamazaki:2001ue}, where black holes with Born-Infeld electrodynamics were obtained. In \cite{HosseinHendi:2017soi} Born-Infeld-Dilaton black holes where also found, discussing also the conserved charges and the thermodynamics. Dilaton black holes with power-Maxwell electrodynamics were also discussed \cite{powermax}.

 In this work we present novel rotating BTZ-like solutions in the Einstein-Maxwell-Dilaton theory. This work can be considered as a generalization of
\cite{Chan:1996rd, Chan:1994qa, Chan:1995wj}. In our work we did not fix the form of the scalar field but it is determined by the field equations. Also the scalar potential is also determined  by the field equations. We followed this approach because the choice of different reasonable potentials leaded to a system of  equations which turned out to be untractable. As a result we followed a bottom up approach and found a theory in which the field equations can be solved analytically determining also the scalar potential.  Solutions for given potentials may still be found but a numerical approach should be followed.

The spacetime is parametrized by six constants, $M,Q,l,\alpha,b,J$, being the black hole mass, electric charge, AdS scale, coupling constant between the Maxwell field and the dilaton, scalar length scale and angular momentum, respectively. The solutions have the interesting feature of reducing to some versions of the BTZ black hole \cite{BTZ1, BTZ2} spacetime by setting the parameters appropriately. For $J=0$ the solutions reduce to the ones obtained at \cite{Dehghani:2017zkm}. To derive these type of solutions we made two assumptions:
\begin{itemize}
\item{The coupling between the dilatonic scalar field and the Maxwell invariant is the dilatonic coupling $f(\phi) = \exp(-2\alpha \phi)$.}
\item{The coupling function $f(\phi)$ modifies the circumference of the black hole according to $f(\phi)R(\phi) =r$} where $R$ is the circumference metric function.
\end{itemize}
Under these two assumptions, we solve the field equations and derive spining solutions. We compare both geometric and thermodynamical aspects of the new solutions with the rotating BTZ case, and we discuss the effects of the dilaton scalar field with the new ingredient,the angular momentum added in this work with respect to the existing literature.

The work is organized as follows: In Section \ref{eq} we set up the theory and derive the field equations. In Section \ref{sol} we solve the field equations to find new rotating dilatonic solutions, while in Section \ref{therm} we discuss thermodynamics and compare the thermodynamical properties of the new solutions with the rotating BTZ case \cite{BTZ1, BTZ2} and \cite{Dehghani:2017zkm}. In Section \ref{a1case} we discuss the particular scenario of $\alpha=1$.
 Finally in Section \ref{conc} we conclude and point out possible extensions of this work that can be investigated in future works.

\section{Basic Field Equations}
\label{eq}

We consider Einstein-Maxwell-Scalar (Dilaton) (EMD) theory in $(2+1)$-dimensions
\begin{equation}
    S=\frac{1}{2}\int d^3x \sqrt{-g} \left[\kappa^{-1}R-2\nabla_\mu \phi \nabla^\mu \phi-V\left(\phi\right)-f\left(\phi\right)F_{\mu\nu}F^{\mu\nu}\right]~, \label{action}
\end{equation}
where $V(\phi)$ is the self-interacting potential of the scalar field while $f(\phi) = \exp (-2\alpha \phi)$ is the dilatonic coupling function to the Maxwell field and we will set  $\kappa=8\pi G=1$. The scalar potential will be left arbitrary and we will determine its form from the field equations.
Varying the action, we can obtain the Einstein equation, Klein-Gordon equation and the electromagnetic field equation
\begin{eqnarray}
&& G_{\mu\nu} \equiv R_{\mu\nu}-\frac{1}{2}R g_{\mu\nu}=T_{\mu\nu}^{\phi}+T_{\mu\nu}^{A}~, \label{Einstein_eq}\\
&& 4\square \phi=V'(\phi)+f'(\phi)F~, \label{KG_eq}\\
&& \partial_\mu \left(\sqrt{-g}f(\phi)F^{\mu\nu}\right)=0~, \label{F_eq}
\end{eqnarray}
where $F \equiv F_{\mu\nu}F^{\mu\nu}$, $F_{\mu\nu}=\partial_\mu A_\nu -\partial_\nu A_\mu$, $A_\mu =\delta_\mu^0 q(r)$, and
\begin{eqnarray}
T_{\mu\nu}^{\phi}&=&2\nabla_\mu\phi \nabla_\nu\phi-g_{\mu\nu}\nabla_\alpha\phi \nabla^\alpha\phi-\frac{1}{2}g_{\mu\nu}V(\phi)~, \\
T_{\mu\nu}^{A}&=&2f(\phi)\left(F_{\mu\sigma}F_\nu{}^\sigma-\frac{1}{4}g_{\mu\nu}F\right)~,
\end{eqnarray}
are stress-energy tensors for the scalar field and the electromagnetic field respectively.

\section{Black Hole Solutions}

\label{sol}
Assuming the rotationally symmetric metric ansatz
\begin{equation}
    ds^2=-B(r)dt^2+\frac{dr^2}{B(r)}+R(r)^2\left(d\theta + u(r) dt\right)^2~, \label{metric1}
\end{equation}
and inserting (\Ref{metric1}) to the field equations, we obtain five independent equations with six unknown functions $B(r),q(r),u(r),R(r),\phi(r),V(r)$, therefore one of the unknowns has to be fixed by hand. By looking at the electromagnetic equation, we can see that it yields a first integral
\begin{equation} f(r)=\frac{Q_0}{R(r) q'(r)}~, \label{f} \end{equation}
where $Q_0$ is the charge parameter. It is clear that the above equation relates the dilatonic coupling with the electric potential and the area of the black hole.
To solve the field equations, as discussed in the Introduction, we will assume that the functions $R(r)$ and $f(r)$ satisfy $f(r)R(r)/r =1$. This particular assumption is also needed in order to make our results comparable with the non-rotating case \cite{Dehghani:2017zkm}. Moreover, the condition $f(r)R(r)/r =1$ can be solved as a result by considering the electric potential $q(r)= Q_0\ln r$ which is reasonable for a $(2+1)-$dimensional theory. Then from (\ref{f}) we can obtain the electric potential and the $R(r)$ function
\begin{eqnarray}
q(r) &=& Q_0 \ln\left(\frac{r}{l}\right)~, \label{electricpotential}\\
R(r) &=& r \exp (2 \alpha  \phi (r))~,\label{r}
\end{eqnarray}
where $l$ is the cosmological scale. We find that it is more convenient to express the solution in terms of a parameter $x$ which is related to the dilatonic  parameter $\alpha$ via
\begin{equation} \alpha =\sqrt{\frac{x}{2 (1-x)}}~,\end{equation}
where $x$ is restricted to take values between $0 \le x <1$.
Given the above configurations, we can obtain the angular shift function $u(r)$ and the scalar field $\phi(r)$
\begin{eqnarray}
&&u(r) = \frac{J_0 r^{3 x-2}}{3 x-2} + c_u~,\\
&&\phi(r) = (1-x) \sqrt{\frac{x}{2-2 x}} \ln \left(\frac{b}{r}\right)~, \label{scalarfield}
\end{eqnarray}
where $J_0$ is the parameter for angular momentum of the black hole, $c_u$ is an integration constant, which in order to remove global rotations of the co-ordinate system we will set it to zero and $b$ is the scalar charge. Also, finiteness of the angular velocity at large distances constraint the coupling constant x as $0\le x<2/3$. We can calculate the lapse function $B(r)$ as
\begin{equation}
B(r) = J_0^2\frac{ b^{2 x} r^{4 x-2}}{(2-3 x)^2}+\frac{2 r^2 }{l^2 \left(3 x^2-5 x+2\right)}\left(\frac{b}{r}\right)^{2 x}+\frac{4 Q_0^2  }{3 x-2}\left(\frac{b}{r}\right)^{-x} \ln \left(\frac{r}{l}\right)-M_0 r^x~,
\end{equation}
which breaks down for $x=2/3$, which corresponds to $\alpha=1$ (low energy strings), $M_0$ is the mass parameter of the black hole and corresponds to the BTZ black hole mass, while the potential supporting this solution is found to be
\begin{equation}
V(r) =-\frac{2 b^{2 x} r^{-2 x}}{l^2} + J_0^2\frac{ x b^{2 x} r^{4 x-4}}{4-6 x}+Q_0^2\frac{2  x b^{-x} r^{x-2}}{2-3 x}~,
\end{equation}
which vanishes at spatial infinity and is independent of the mass $M_0$, while as a function of $\phi$ reads
\begin{equation}
V(\phi) =-\frac{2}{l^2} \exp\left({\frac{2 \sqrt{2} x \phi }{\sqrt{(1-x) x}}}\right)-J_0^2\frac{ x b^{6 x-4}}{2 (3 x-2)} \exp\left(\frac{4 \sqrt{2-2 x} \phi }{\sqrt{x}}\right)+Q_0^2\frac{2  x }{b^2 (2-3 x)}\exp\left({\frac{\sqrt{2-2 x} (x-2) \phi }{(x-1) \sqrt{x}}}\right)~.
\end{equation}
We can see that the resultant potential is $V(\phi) \sim \sum_{i=1}^{3}a_{i}e^{\lambda_{i}\phi}$ i.e a linear combination of Liouville-type potentials, where $\lambda_{i}$ are the co-efficients of the exponents and $a_{i}$ are the coupling constants.
For $\alpha \to 0$ (i.e $x \to 0$), the solution yields the BTZ black hole \cite{BTZ1}
\begin{eqnarray}
B(r) &=& \frac{J^2}{4 r^2}+\frac{r^2}{l^2}-2 Q^2 \ln \left(\frac{r}{l}\right)-M_0~,\\
u(r) &=& -\frac{J}{2r^2}~,\\
V(r) &=& -\frac{2}{l^2}~,
\end{eqnarray}
and the electric potential is given by (\Ref{electricpotential}), while all other functions vanish due to their dependence on $\alpha$.
The obtained scalar field has the standard $(2+1)$-dimensional dilaton form \cite{Dehghani:2017zkm} and diverges at both  the origin and spatial infinity. Despite the fact that the scalar field diverges at large distances, the action remains finite since
\begin{equation} \lim_{r \to \infty} \Big(\nabla_{\sigma}\phi\nabla^{\sigma}\phi\Big) \to 0~,\end{equation}
while  all the other quantities of the action (\Ref{action}) also remain finite for large $r$.

\begin{figure}
\centering
\includegraphics[width=.40\textwidth]{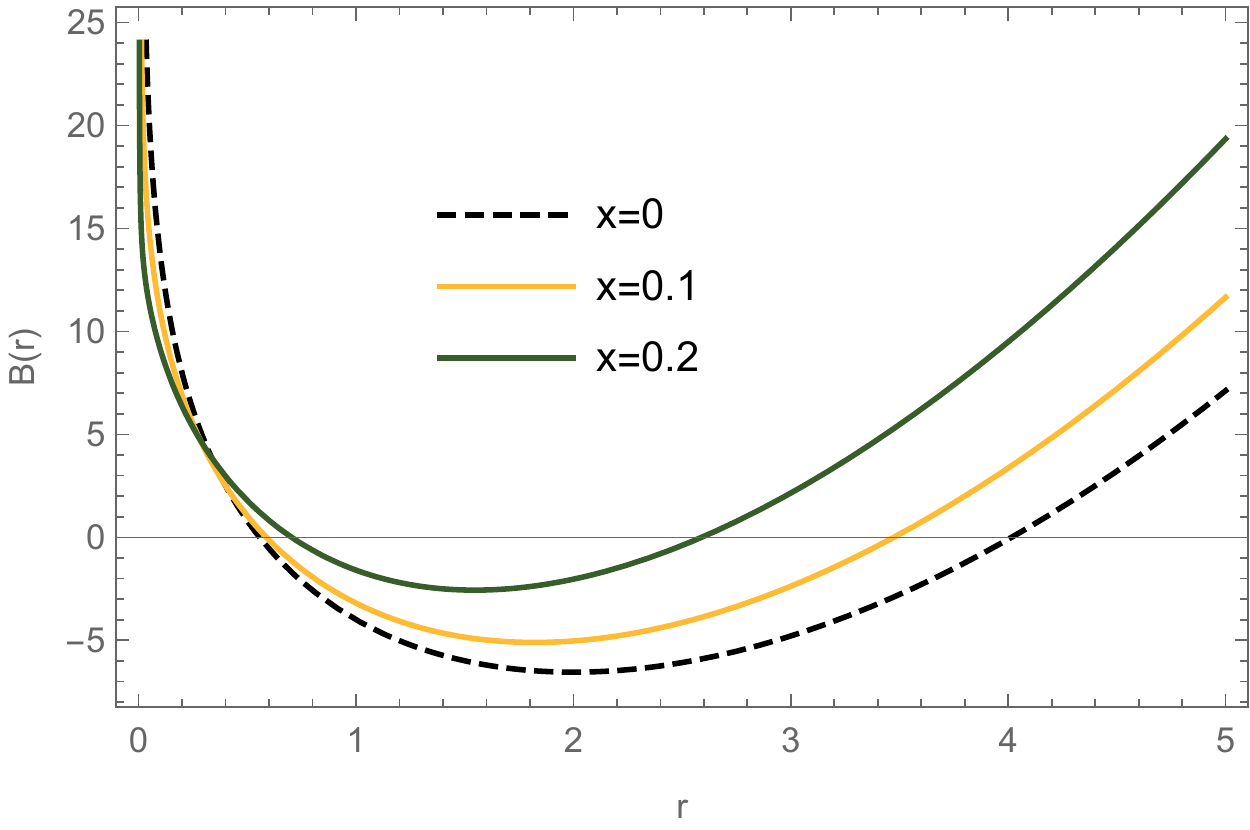}
\includegraphics[width=.40\textwidth]{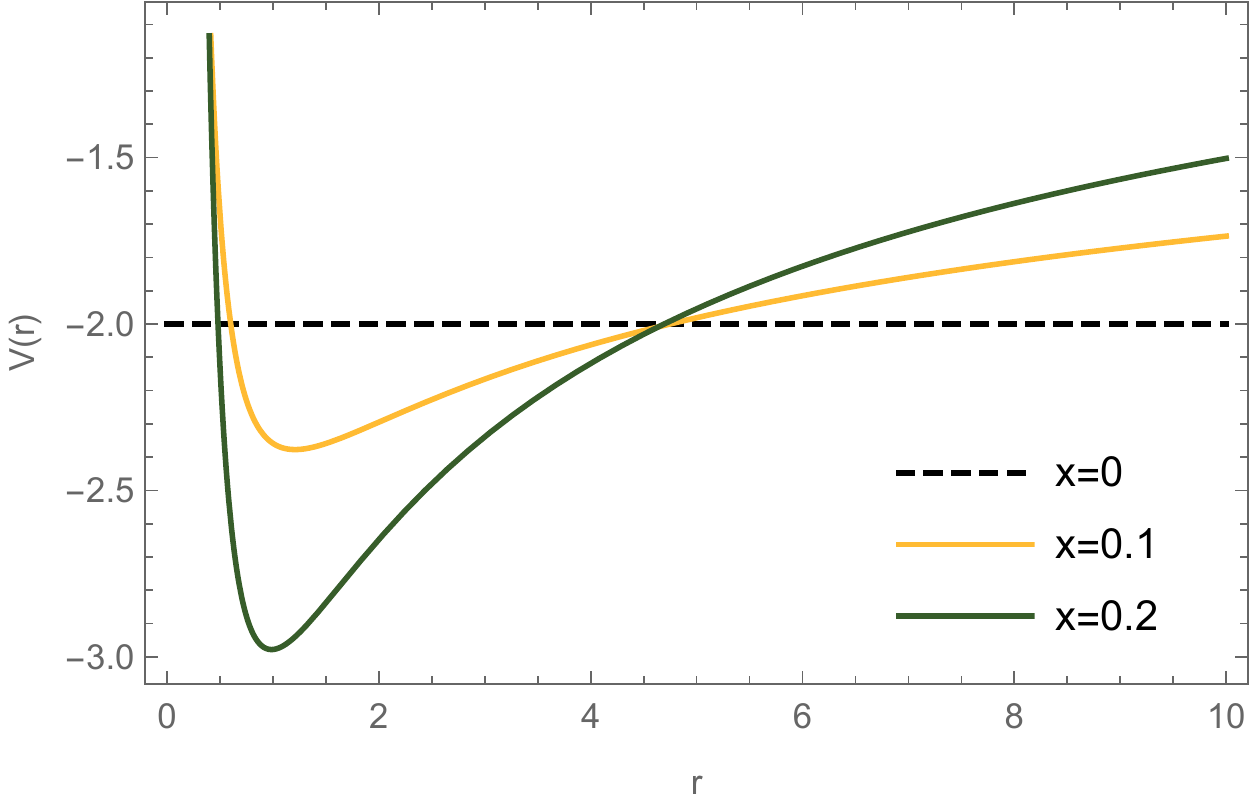}
\includegraphics[width=.40\textwidth]{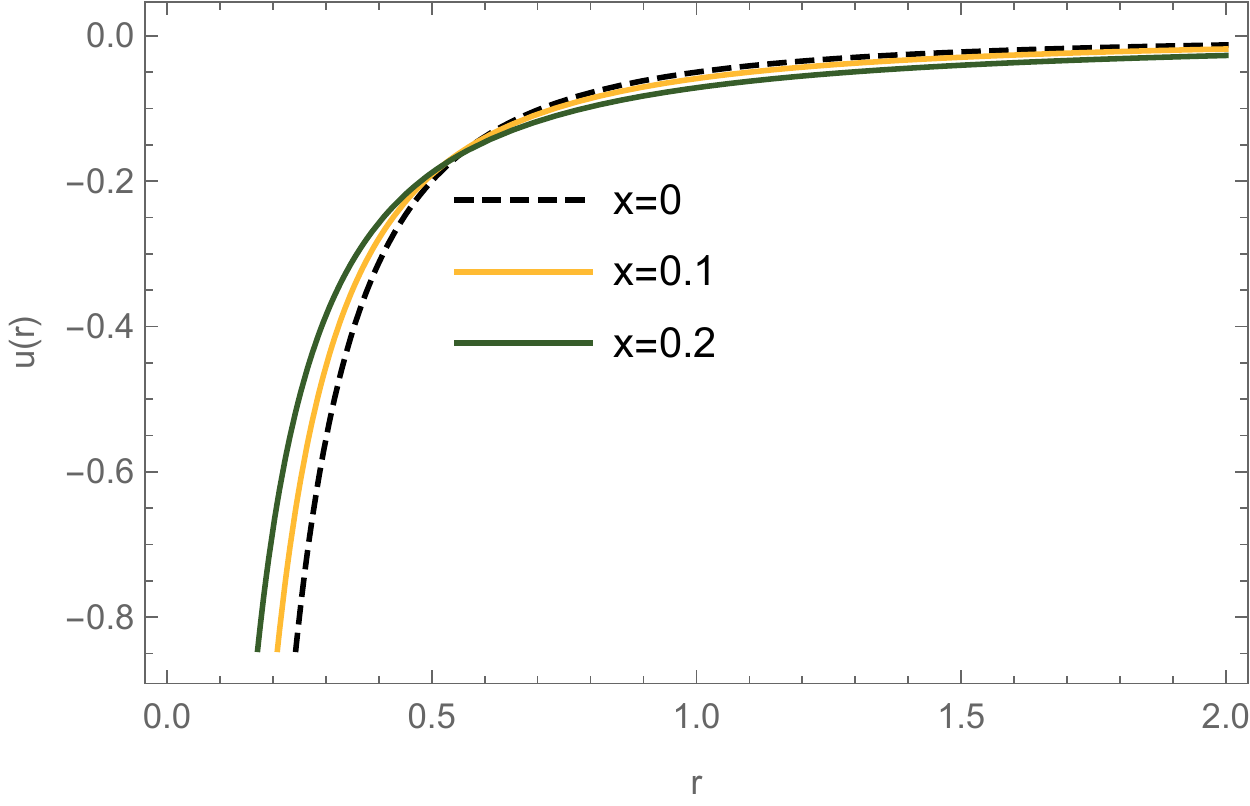}
\includegraphics[width=.40\textwidth]{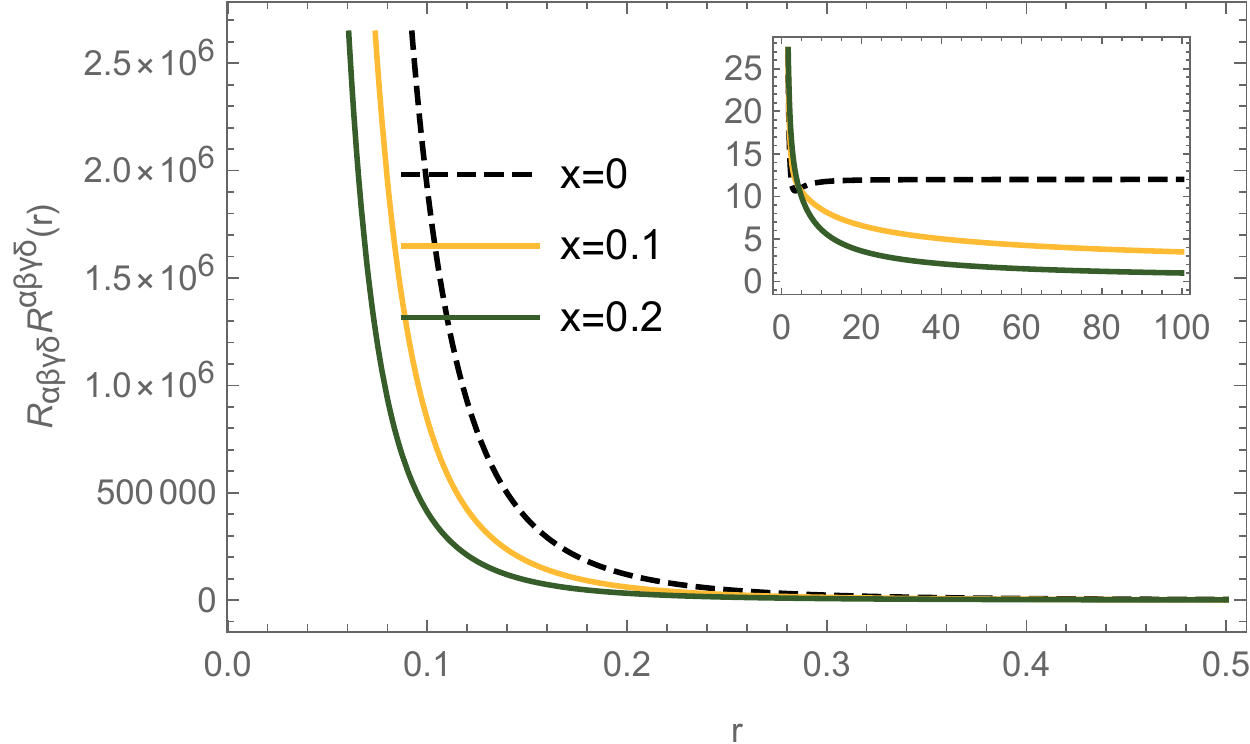}
\includegraphics[width=.40\textwidth]{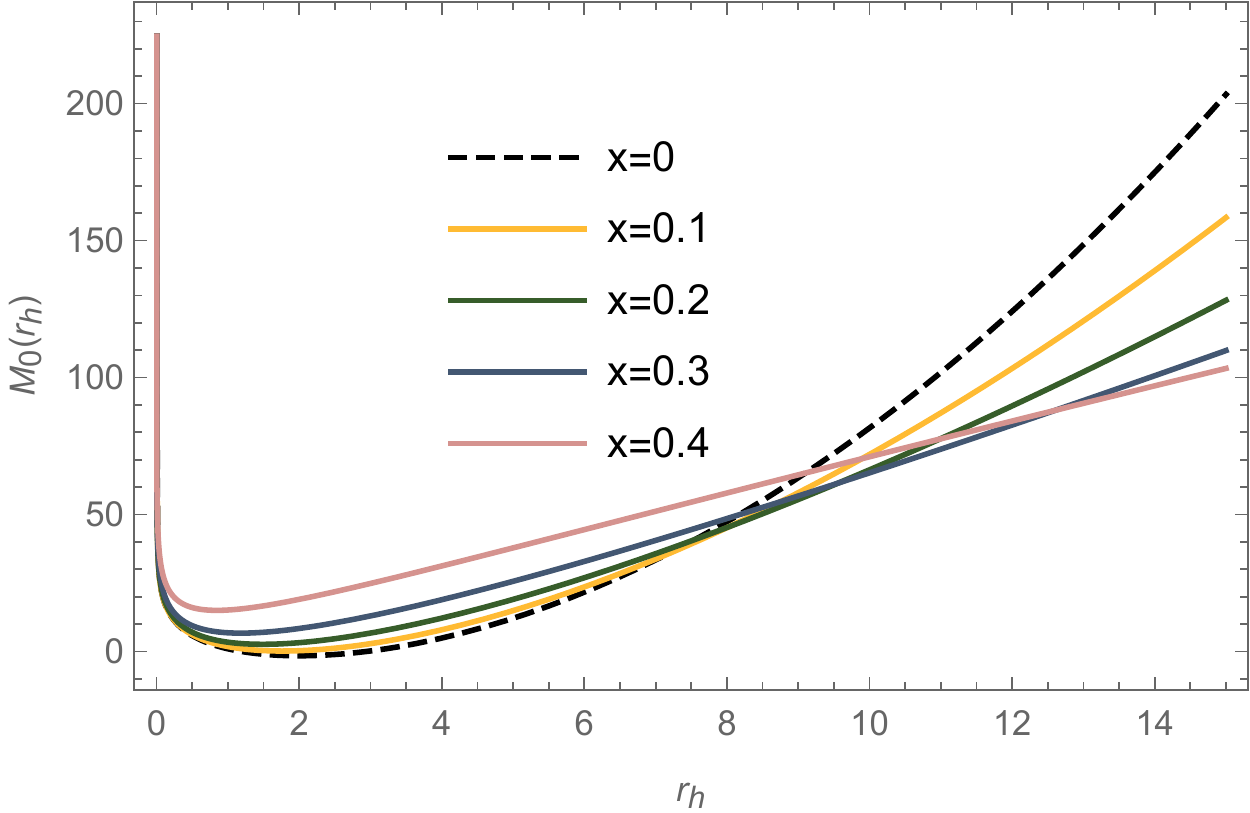}
\caption{The functions $B(r)$, $V(r)$. $u(r)$ and the Kretschmann scalar versus $r$, while changing the model parameter $\alpha$ through $x$, having set $l=1, b =M_0 =5, J_0=0.1, Q_0=2$. We also plot $M_0(r_h)$ as a function of $r_h$.} \label{met_V}
\end{figure}

In Fig. \Ref{met_V} we plot the metric function $B(r)$, the potential $V(r)$, the angular shift function $u(r)$ and the Kretschmann scalar $R_{\alpha\beta\gamma\delta}R^{\alpha\beta\gamma\delta}$ along with the $x=0$ case which corresponds to the charged and rotating BTZ black hole and also the geometric mass $M_0$ with respect to the horizon. From these figures, we can see that the metric function develops a smaller radius for the event horizon of the black hole, the potential $V(r)$ for the dilaton case $x\neq0$ goes to zero at large distances, indicating that spacetime is not asymptotically AdS, as in the BTZ case. It also develops a potential well between the inner and event horizon of the black hole. The Kretschmann scalar diverges at the origin $r=0$, indicating a physical singularity (the singularity in the BTZ case comes from the electric charge $Q_0$, if $Q_0=0$, the BTZ spacetime is completely regular everywhere) and asymptotically goes to zero for large $r$, while it is related to the AdS radius for the BTZ case. The black hole mass has a minima and grows with the increase of the horizon, however the BTZ case possesses a higher mass for large values of $r_h$.
\begin{figure}
\centering
\includegraphics[width=.40\textwidth]{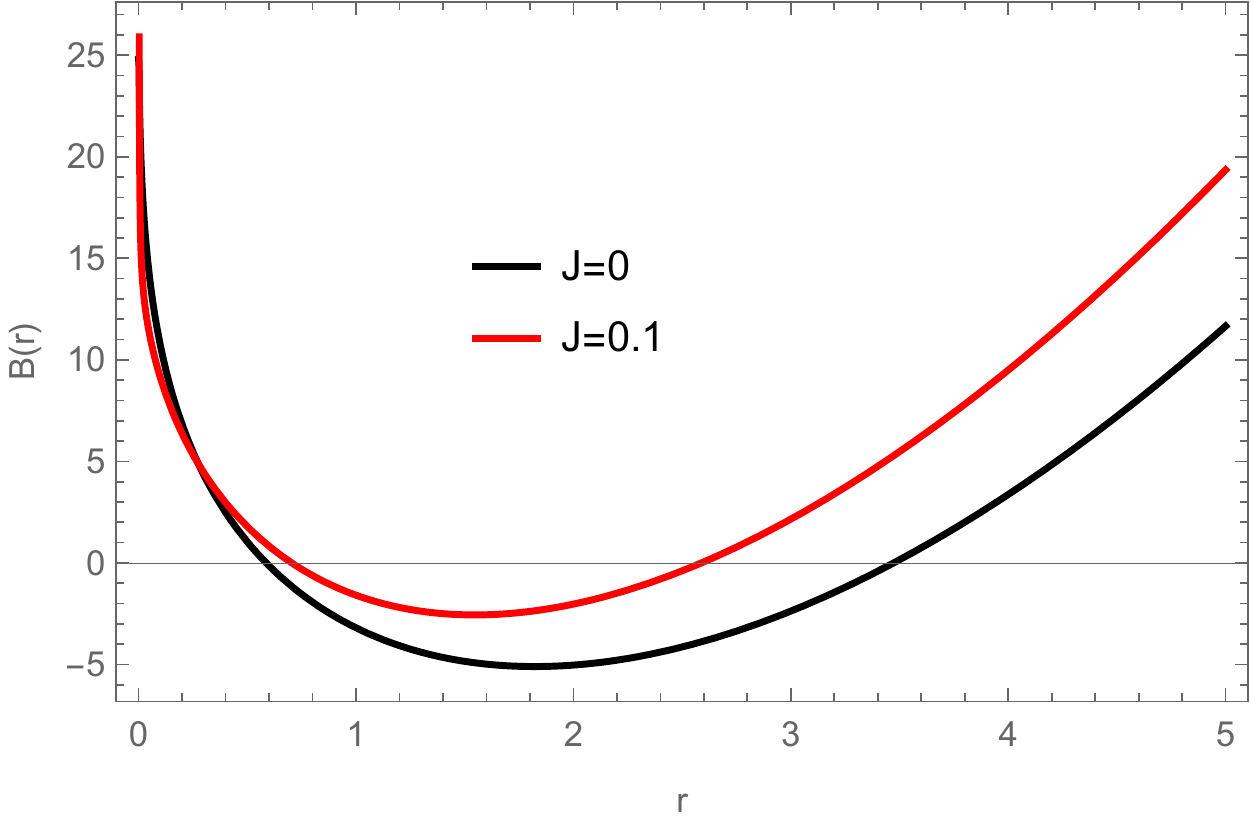}
\includegraphics[width=.40\textwidth]{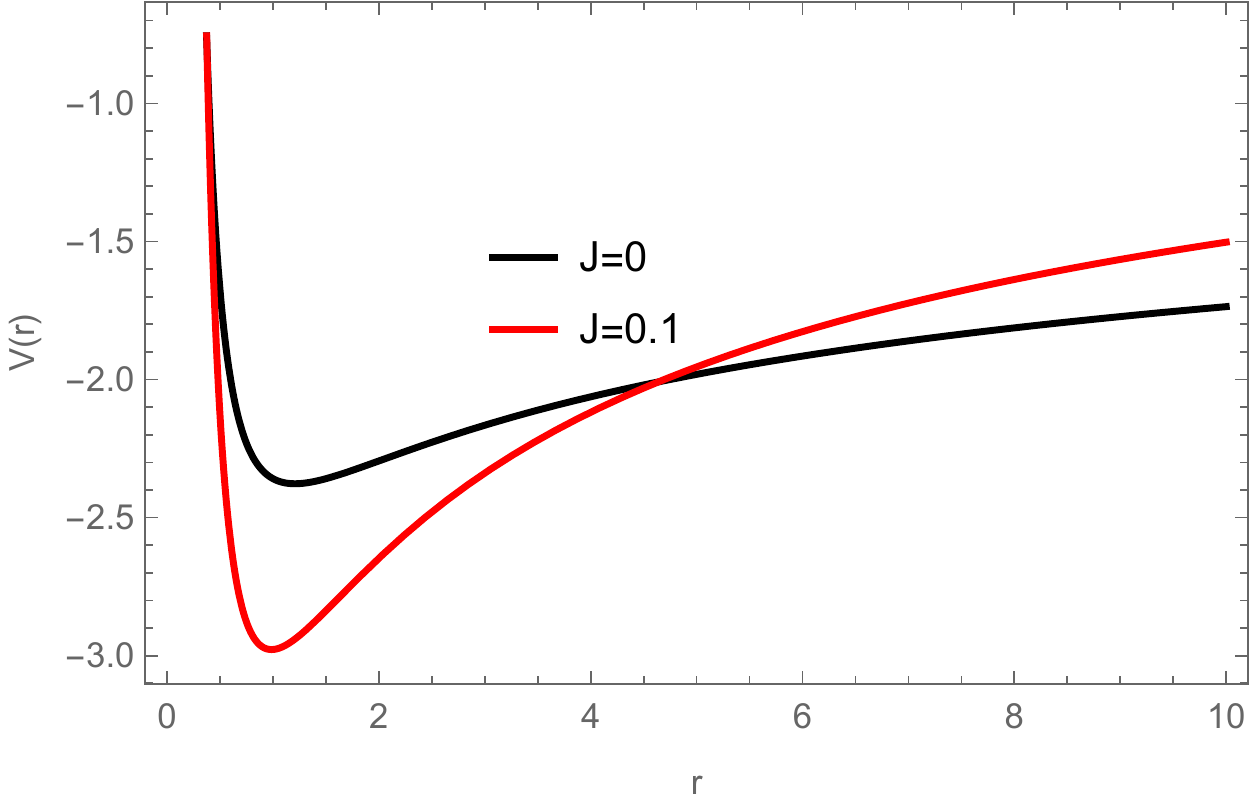}
\caption{The metric function $B(r)$ and the potential $V(r)$ for $J_0=0$ and $J_0=0.1$ having set $x=0.1,l=1,Q_0=2,M_0=b=5$.} \label{bvj}
\end{figure}

In Fig. \Ref{bvj} we plot $B(r)$ and $V(r)$ for the static and rotating dilatonic black holes respectively in order to see the effect of the angular momentum on the resulting physics. We can see that the rotating solution possesses a smaller event horizon, while the potential develops a deeper potential well inside the event horizon of the black hole.

\section{Themrodynamics and Conserved Charges}

\label{therm}
In this Section we will discuss the thermodynamics of the obtained black hole solution. We first discuss  the Hawking temperature. To compute temperature we need to go to Euclidean space, which further specifies the periodicity of coordinates $\tau,\theta$ as $\tau\to \tau +\beta~, \theta \to \theta - \Omega\beta$, where $\Omega$ is the angular velocity of the horizon (see the discussion below) and $\beta$ is related to the inverse of temperature which is given by
\begin{equation}
T(r_h) = \cfrac{1}{\beta} = \frac{B'(r_h)}{4\pi}~,
\end{equation}
where $r_h$ represents the position of the outer horizon, while its explicit expression yields
\begin{equation}
T(r_h) = \frac{b^{-x} r_h^{-2 x-3} \left(b^{3 x} \left(J_0^2 l^2 (x-1) r_h^{6 x}+(4-6 x) r_h^4\right)+4 l^2 Q_0^2 (x-1) r_h^{3 x+2}\right)}{4 \pi  l^2 (x-1) (3 x-2)}~,
\end{equation}
where we have already substituted $M_0$ from the $B(r_h)=0$ condition of the black hole horizon. Its expansion for large coupling constant $\alpha>>1,~x\to1$ yields
\begin{equation} T(r_h) = -\frac{b^2}{2 (x-1) \left(\pi  l^2 r_h\right)} +\mathcal{O}\left((x-1)^0\right)~.\end{equation}
For $x\to0$, we obtain
\begin{equation}
T(r_h) = -\frac{J_0^2}{8 \pi  r_h^3}+\frac{r_h}{2 \pi  l^2}-\frac{Q_0^2}{2 \pi  r_h}~, \label{tempbtz}\end{equation}
which is the BTZ black hole temperature. The zero point of the temperature corresponds to $B'(r_h)=0$, i.e. extremal black holes, and can be obtained analytically as
\begin{equation} r_\text{extremal} = (12 x-8)^{\frac{1}{3 x-2}} b^{\frac{3 x}{3 x-2}} \left(4 l^2 Q_0^2 (x-1)-2 \sqrt{2} \sqrt{l^2 (x-1) \left(J_0^2 (3 x-2) b^{6 x}+2 l^2 Q_0^4 (x-1)\right)}\right)^{\frac{1}{2-3 x}}~,\end{equation}
while the extremal horizon of BTZ case is given by
\begin{equation} r_\text{extremal} = \frac{1}{\sqrt{2}}\sqrt{l \left(\sqrt{J_0^2+l^2 Q_0^4}+l Q_0^2\right)}~.\end{equation}
The entropy will be given by the Bekenstein-Hawking area law, since the theory (\Ref{action}) is a minimally coupled to gravity theory and higher order invariants are absent, so the only term that will contribute to Wald entropy \cite{Wald:1993nt} is the pure Ricci scalar term which yields the entropy as
\begin{equation} S= \frac{\mathcal{A}}{4G} =2\pi \mathcal{A}= 4 \pi^2 b^x r_h^{1-x}~, \end{equation}
where $\mathcal{A} = 2\pi b^x r_h^{1-x}$ is the area of black hole.
It can be expanded near the no coupling case $x=0$ ($\alpha=0$)
\begin{equation}
    S=\frac{4 \pi ^2 r_h}{\kappa }+\frac{4 \pi ^2 r_h }{\kappa}\ln{\left(\frac{b}{r_h}\right)}x+\mathcal{O}(x^2)~, \label{S0}
\end{equation}
while for large coupling constant $\alpha \gg 1,~x\to1$ we have
\begin{equation} S= \frac{\pi  b}{2} + \frac{1}{2} \pi  b (x-1) \ln{\left(\frac{b}{r_h}\right)}+\mathcal{O}\left((x-1)^2\right) ~.\end{equation}
The Hawking temperature of the black hole can also be expanded near the $x=0$ case
\begin{equation}
    T=\frac{4 \pi ^2 r_h^2 \left(r_h^2-4\right)-1}{8 \pi ^3 r_h^3}+\frac{x \left(8 \pi ^2 \left(r_h^2-6\right) r_h^2-8 \left(2 \pi ^2 \left(r_h^2+2\right) r_h^2+1\right) \ln (r_h)-3\right)}{16 \pi ^3 r_h^3}+\mathcal{O}\left(x^2\right)~.
\end{equation}

\begin{figure}
\centering
 \includegraphics[width=.40\textwidth]{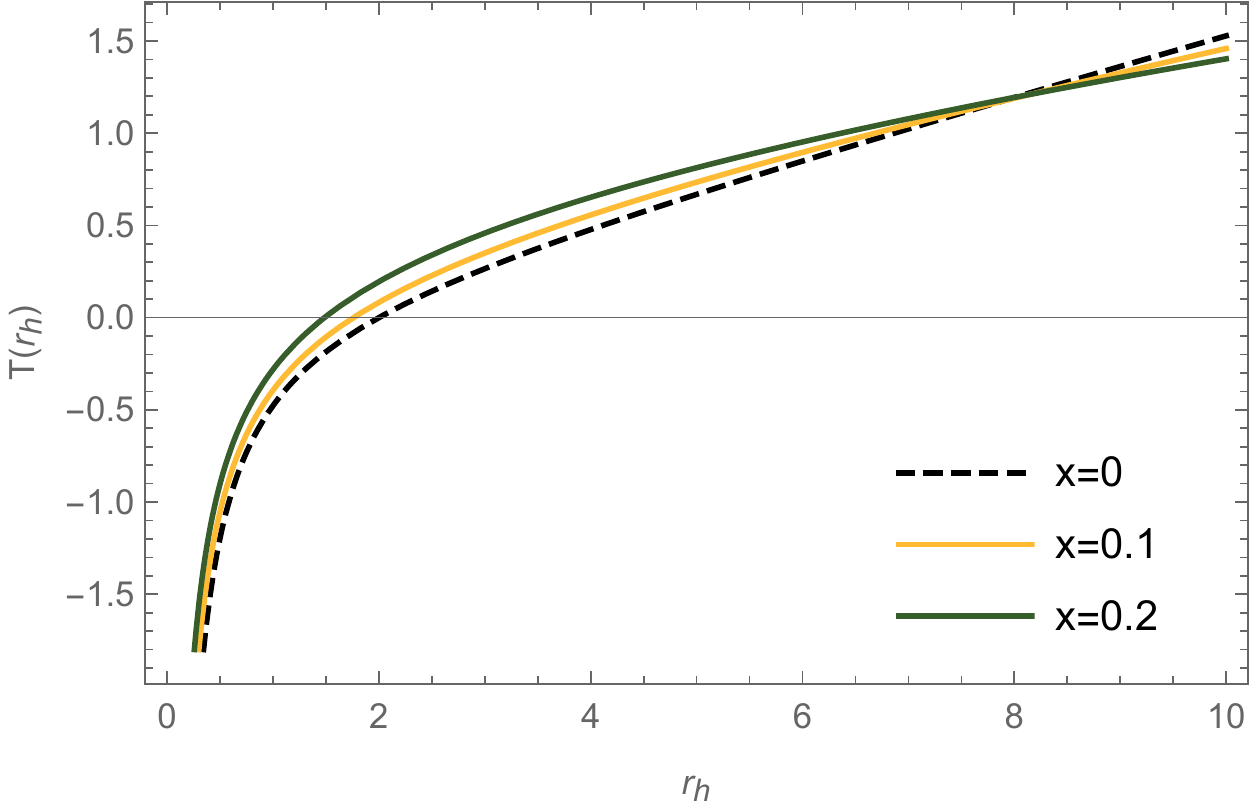}
 \includegraphics[width=.40\textwidth]{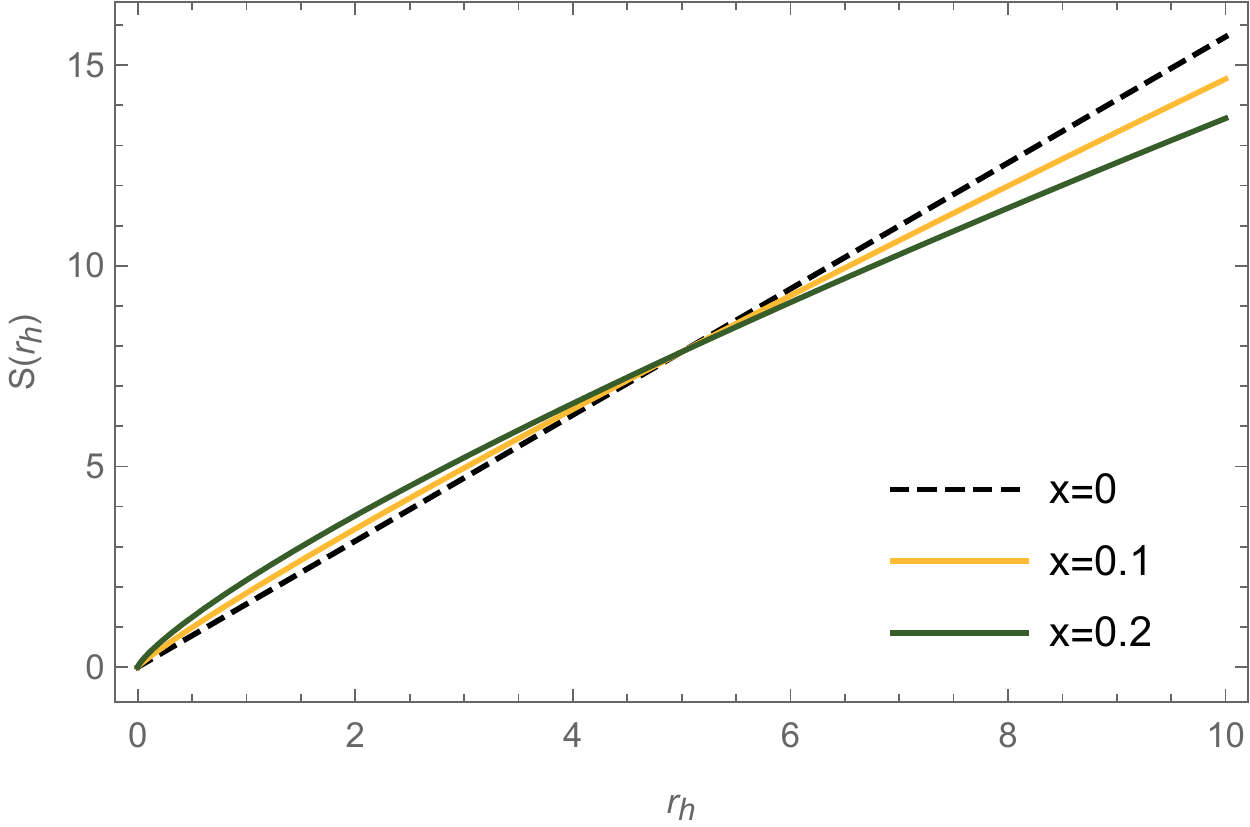}
\caption{The temperature and the entropy versus $r_h$, for different valus of $x$, having set $l=1,J_0=0.1,Q_0=2,M_0=b=5$.} \label{temp_entro}
\end{figure}

We can see that for large coupling constant, the entropy takes a constant value and does not depend at zeroth order on the horizon radius as should be expected. This is another clue that large values of the coupling constant shoulb be excluded, besides the one coming from the finiteness of the angular velocity function as we already discussed.
For $x\to 0$ we recover the BTZ black hole entropy and temperature (\ref{tempbtz}), while we can see that the scalar length scale $b$ affects the entropy. We present plots of the temperature and the entropy for some values of $x$ alongside $x=0$ which corresponds to the BTZ case in order to make comparisons in Fig. \Ref{temp_entro}.

From Fig. \Ref{temp_entro} we can see that the temperature and the entropy of the rotating dilatonic black holes do not differ qualitatively from the corresponding quantities of the BTZ black hole spacetime. The entropy is always positive and the dilatonic black holes may have higher entropy than the BTZ black hole, depending on the value of the scalar length scale $b$. The expression (\ref{S0}) and the Fig. \Ref{temp_entro} both show that for $r_h<b$, the dilatonic black holes are thermodynamically preferred, elsewise the BTZ case is preferred.

To compute the physical electric charge we will use \cite{Dehghani:2017zkm}

\begin{equation}
Q = \frac{1}{4\pi} \int \sqrt{-g}  \frac{\partial \mathcal{L}_{EM}}{\partial F}F_{\mu\nu} n^{\mu} v^{\nu} d\Omega= \frac{1}{4\pi} \int dS^{\mu}f(\phi)\nabla_{\mu}q(r) = - \frac{Q_0}{2}~, \end{equation}
where $\mathcal{L}_{EM}$ denotes the electromagnetic part of the Lagrangian of our theory (\Ref{action}) and $n^{\mu},~ v^{\nu}$ are the unit spacelike and timelike normal vectors to the hypersurface of radius $r$ \cite{Dehghani:2017zkm}. We can see that the dilatonic coupling does not affect the conserved electric charge which is expected since we assumed that $f(r)R(r)=r$, so that the effect of the dilatonic coupling has the exact counter-effect on the circumference of the black hole. It also coincides with the non-rotating case \cite{Dehghani:2017zkm} as expected.

To compute the mass of the black hole we will use the quasilocal method \cite{Brown:1992br,Brown:1994gs}. The metric (\ref{metric1}) can be transformed to a more general form
\begin{equation}
    ds^2=-N^2(R)dt^2+\frac{dR^2}{H^2(R)}+R^2\left[d\theta+ V^\theta(R) dt\right]^2~, \label{MetricR}
\end{equation}
with
\begin{eqnarray}
&& N^2(R)=B\left(b^{\frac{x}{x-1}} R^{\frac{1}{1-x}}\right)~,\\
&& H^2(R)=B\left(b^{\frac{x}{x-1}} R^{\frac{1}{1-x}}\right)(1-x)^2 b^{\frac{2 x}{1-x}} R^{-\frac{2 x}{1-x}}~,\\
&& V^\theta(R)=u\left(b^{\frac{x}{x-1}} R^{\frac{1}{1-x}}\right)~.
\end{eqnarray}

For a $D$-dimensional spacetime manifold $\mathcal{M}$, which is topologically the product of a spacelike hypersurface and a real line interval $\Sigma \times I$, the total quasi-local energy is defined as \cite{Brown:1992br,Brown:1994gs}
\begin{equation}
    E=\int_B d^{D-2}x \sqrt{\sigma}\varepsilon~,
\end{equation}
where $B\equiv \partial \Sigma$ is the $(D-2)$-dimensional boundary, $\sigma$ is the determinant of the induced metric $\sigma_{ab}$ on $B$, and $\varepsilon$ is the energy density.

The boundary $\partial\mathcal{M}$ consists of initial and final spacelike hypersurfaces $t'$ and $t''$ respectively, and a timelike hypersurface $\mathcal{T}=B\times I$ joining these. The $(D-1)$-metric $\gamma_{ij}$ on $\mathcal{T}$ can be written according to the Arnowitt, Deser, Misner (ADM) decomposition as
\begin{equation}
    \gamma_{ij}dx^i dx^j=-N^2 dt^2+\sigma_{ab}\left(dx^a+V^a dt\right)\left(dx^b+V^b dt\right)~.
\end{equation}
The conserved charge associated with a Killing vector field $\xi^i$ is defined as
\begin{equation}
    Q_\xi=\int_B d^{D-2}x\sqrt{\sigma}\left(\varepsilon u^i+j^i\right)\xi_i~,
\end{equation}
where $u^i$ is the unit normal to spacelike hypersurfaces $t'$ or $t''$, and
$j^i$ is the momentum density.

If the system contains a rotational symmetry given by a Killing vector field $\zeta^i=(\partial/\partial\theta)^i$ on $\mathcal{T}$, and the $(D-2)$-surface $B$ is chosen to contain the orbits of $\zeta^i$, then the angular momentum can be expressed as
\begin{equation}
    J=Q_\zeta=\int_B d^{D-2}x \sqrt{\sigma}j_i \zeta^i~.
\end{equation}

Also, a conserved mass can be given by
\begin{equation}
    M=-Q_\xi=-\int_B d^{D-2}x\sqrt{\sigma}\left(\varepsilon u_i+j_i\right)\xi^i~,
\end{equation}
where $\xi^i=N u^i+V^a (\partial/\partial x^a)^i$ is the associated timelike Killing vector field.

The energy density $\varepsilon$ and quasi-local energy $E$ can be calculated in metric form (\ref{MetricR})
\begin{eqnarray}
    && \varepsilon(R)=k(R)-\varepsilon_0(R)~, \\
    && E(R)=2\pi R \  \varepsilon (R)=2\pi R k(R)  -2\pi R \varepsilon_0(R)~,
\end{eqnarray}
where $k(R)=-H(R)/R$ is the trace of the extrinsic curvature of $B$. To make the total quasi-local energy finite, the condition $0 \le x <2/3$ is required and $\varepsilon_0$ is chosen to be
\begin{equation}
    \varepsilon_0(R)=-\frac{1}{l}\sqrt{\frac{2(1-x)}{2-3 x}} \left(\frac{b}{R}\right)^\frac{x}{1-x}~.
\end{equation}

The only nonzero components of the momentum $P^{ij}$ and the momentum density $j_i$ are
\begin{eqnarray}
    && P^{r\theta}=P^{\theta r}=-\frac{R H(R)}{4   N(R)}\frac{dV^\theta(R)}{dR}=-\frac{ b^{3 x}J_0}{4   R^2}~,\\
    && j_\theta=\frac{ b^{3 x}J_0}{2   R}~,
\end{eqnarray}
then we can calculate the total angular momentum associated with the Killing vector field $\zeta^i=(\partial/\partial\theta)^i$ as
\begin{equation}
    J=2 \pi R j_\theta=\pi b^{3 x} J_0~. \label{j1}
\end{equation}

Finally the conserved mass can be obtained by taking the limit at the space infinity
\begin{equation}
    M=\lim_{R \to \infty} \left( N(R)E(R)-2\pi R j_\theta \xi^\theta\right)=\pi (1-x) b^x M_0~,\label{m1}
\end{equation}
which can reduce to the ADM mass $M_0$ of the BTZ case when $x=0$.
We can see that the scalar length scale $b$ enters in both conserved mass and the conserved angular momentum, therefore the scalar hair of the solution is of the  secondary type, which is not given by a unique Gauss law but related to the black hole conserved quantities. Using the relations (\ref{j1}) and (\ref{m1}) we get
\begin{equation}
    \frac{J}{M}=\frac{b^{2x}J_0}{(1-x)M_0}~.
\end{equation}

\begin{figure}
\centering
\includegraphics[width=.45\textwidth]{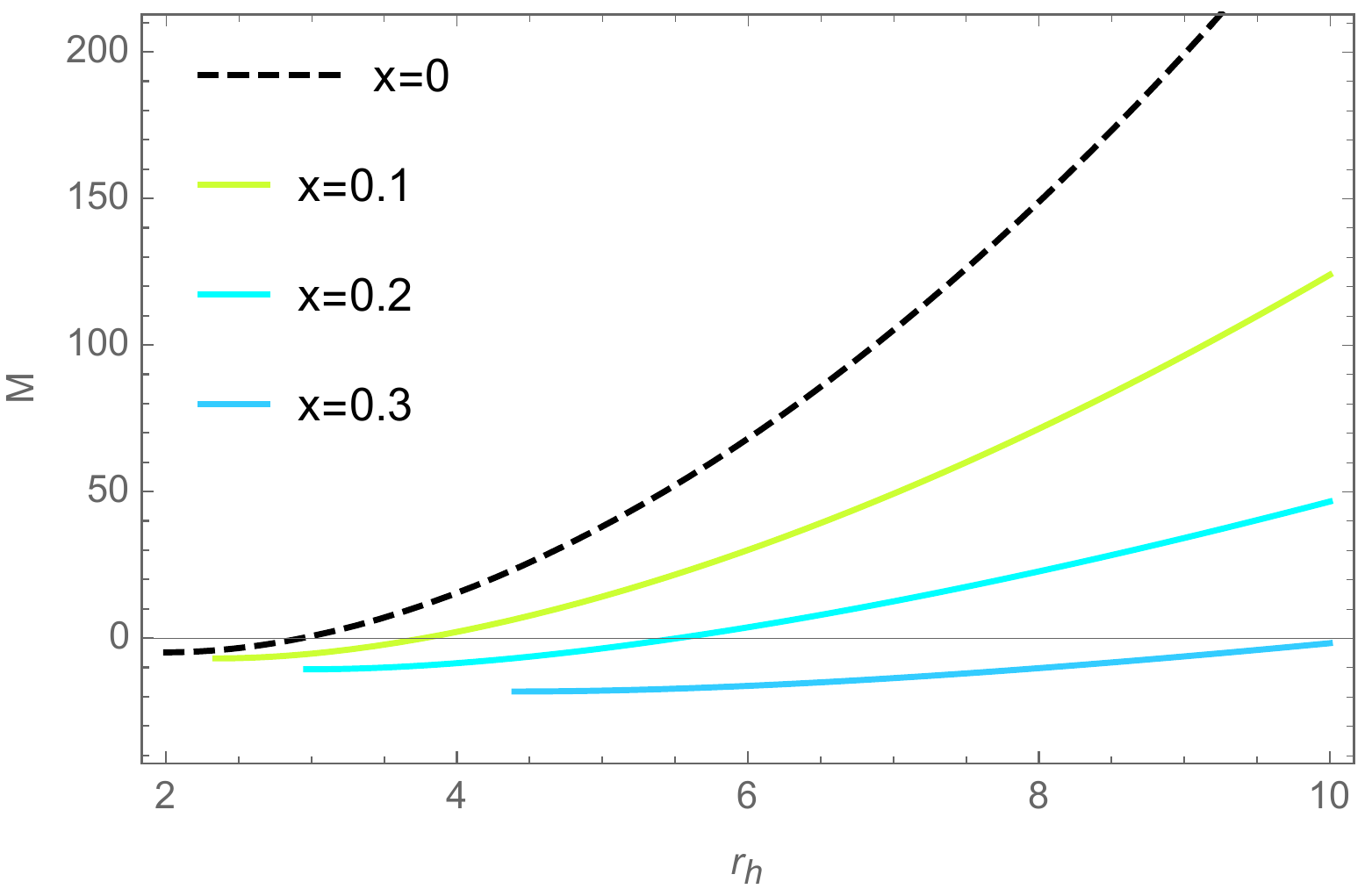}
\caption{The conserved mass $M$ is plotted as a function of the event horizon $r_h$ with various values of $x$, where we have set $Q=J=b=l=1$.}
\label{M_rh}
\end{figure}

Substituting the conserved mass $M$, conserved angular momentum $J$ and conserved charge $Q=-Q_0/2$ into the metric and taking $B(r_h)=0$, we obtain
\begin{equation}
    M\left(r_h,Q,J\right)=\frac{\pi  (x-1) b^x r_h^{-x} }{\kappa }\left(-\frac{\kappa ^2 J^2 b^{-4 x} r_h^{4 x-2}}{\pi ^2 (2-3 x)^2}-\frac{2 b^{2 x} r_h^{2-2 x}}{l^2 \left(3 x^2-5 x+2\right)}+\frac{16 Q^2 b^{-x} r_h^x }{2-3 x}\ln \left(\frac{r_h}{l}\right)\right)~.
\end{equation}
In Fig. \ref{M_rh}, we plot the figure of conserved mass $M$ as a function of the event horizon $r_h$ with various values of $x$. We can see that the conserved mass $M$ always grows with the increase of $r_h$, and its tendency descends as the coupling constant $x$ (or $\alpha$) rises, which means that in 3D EMD theory the black holes of the same size contain less energy.

The partial derivative of the conserved mass $M$ with respect to entropy $S$
\begin{equation}
    \frac{\partial M}{\partial S}=\frac{\partial M/\partial r_h}{\partial S/\partial r_h}=\frac{B'(r_h)}{4\pi}=T~,
\end{equation}
is exactly the Hawking temperature at the event horizon. The electric potential $\Phi$ at the event horizon can be calculated using the null generator of the horizon $\chi = C_0\partial^{t} + \Omega_h\partial^\theta$ as
\begin{equation}
    \Phi= A_{\mu}\chi^{\mu}\Big|_{\infty} - A_{\mu}\chi^{\mu}\Big|_{r_h} = C_0 Q_0\ln\frac{r_h}{l}~,
\end{equation}
while the derivative of the conserved mass $M$ with respect to $Q$ leads to
\begin{equation}
    \frac{\partial M}{\partial Q}=\frac{32 \pi  Q (x-1) }{  (2-3 x)}\ln \left(\frac{r_h}{l}\right)=\Phi~,
\end{equation}
where we have set $C_0=\frac{32 \pi  (x-1) }{  (2-3 x)}$. However the derivative of the conserved mass $M$ with respect to angular momentum $J$ is proportional to the angular velocity $\Omega_h=-u(r_h)$ at the event horizon with a coefficient depending on the coupling constant
\begin{equation}
    \frac{\partial M}{\partial J}=-\frac{2   J (x-1) b^{-3 x} r_h^{3 x-2}}{\pi  (2-3 x)^2}=\frac{2 (1-x)}{2-3 x} \Omega=\frac{1}{1-\alpha^2} \Omega~,
\end{equation}
which indicates that the first law is modified in EMD theory in $(2+1)$-dimensions
\begin{equation}
    dM=TdS+\Phi dQ+\frac{1}{1-\alpha^2}\Omega dJ~.
\end{equation}

 We should note here, that the conserved charge $Q$ and conserved angular momentum $J$ are parts of the theory under consideration, since they appear in the potential $V(\phi)$. Consequently, these are not pure integration constants that are allowed to vary, but constants of the theory, so varying $Q,J$ is like varying the whole theory. The mass $M$ does not appear in the potential, so the first law should (at least) satisfy $M'(r_h)=T(r_h)S'(r_h)$ and indeed this holds. In conclusion, the anomalous angular momentum term is a consequence of the fact that our theory can only yield solutions with a particular angular momentum. Nevertheless, it is worth noting that in \cite{Chan:1995wj}, the uncharged rotating solution in $(2+1)$-dimensional EMD theory does not satisfy the $dJ$ term of the first law either, i.e. $\partial M/\partial J\neq \Omega$, where $J$ is not a parameter of the theory.

\section{Solution with $\alpha =1$}
\label{a1case}

In this section we will discuss the case of $\alpha=1$. The motivation for this consideration is in the case of $\alpha=1$ the action (\ref{action}) is resulting from a string theory with  a dilatonic field to couple to a gauge field. So it would be interesting to see if there are rotating black hole solutions coming from this string theory setup. We have seen that the general case breaks down for $\alpha=1,~x=2/3$ and for this reason we will set $\alpha=1$ first and then solve the field equations. The solution writes as

\begin{eqnarray}
&&\phi(r) = \frac{1}{3} \ln \left(\frac{b}{r}\right)~,\\
&&R(r) = f(r)^{-1}=b^{2/3} r^{1/3}~,\\
&&u(r)=J_0 \ln \left(\frac{r}{b}\right)~,\\
&&B(r) = b^2 J_0^2 \ln ^2\left(\frac{r}{b}\right)\left(\frac{b}{r}\right)^{-2/3}-6 \left(\frac{r}{b}\right)^{2/3} \ln \left(\frac{r}{l}\right)
   \left(Q^2 \left(\frac{1}{3} \ln \left(\frac{b^2}{l r}\right)+1\right)-\frac{b^2}{l^2}\right)-M_0 r^{2/3}~,\\
&&V(r) = \frac{J_0^2 \left(4 b^2 \ln \left(\frac{b}{r}\right)-3 b^2\right)}{6 b^{2/3} r^{4/3}}-\frac{2 b^{4/3}}{l^2 r^{4/3}}+\frac{4 Q^2 }{3 b^{2/3} r^{4/3}}\ln
   \left(\frac{b}{r}\right)~,\\
&&V(\phi) = \frac{J_0^2 e^{4 \phi } \left(4 b^2 \left(\ln \left(\frac{l}{b}\right)+3 \phi \right)-3 b^2\right)}{6 b^2}+\frac{4 Q^2 e^{4 \phi } \phi
   }{b^2}-\frac{2 e^{4 \phi }}{l^2}~.
\end{eqnarray}

\begin{figure}
\centering
\includegraphics[width=.40\textwidth]{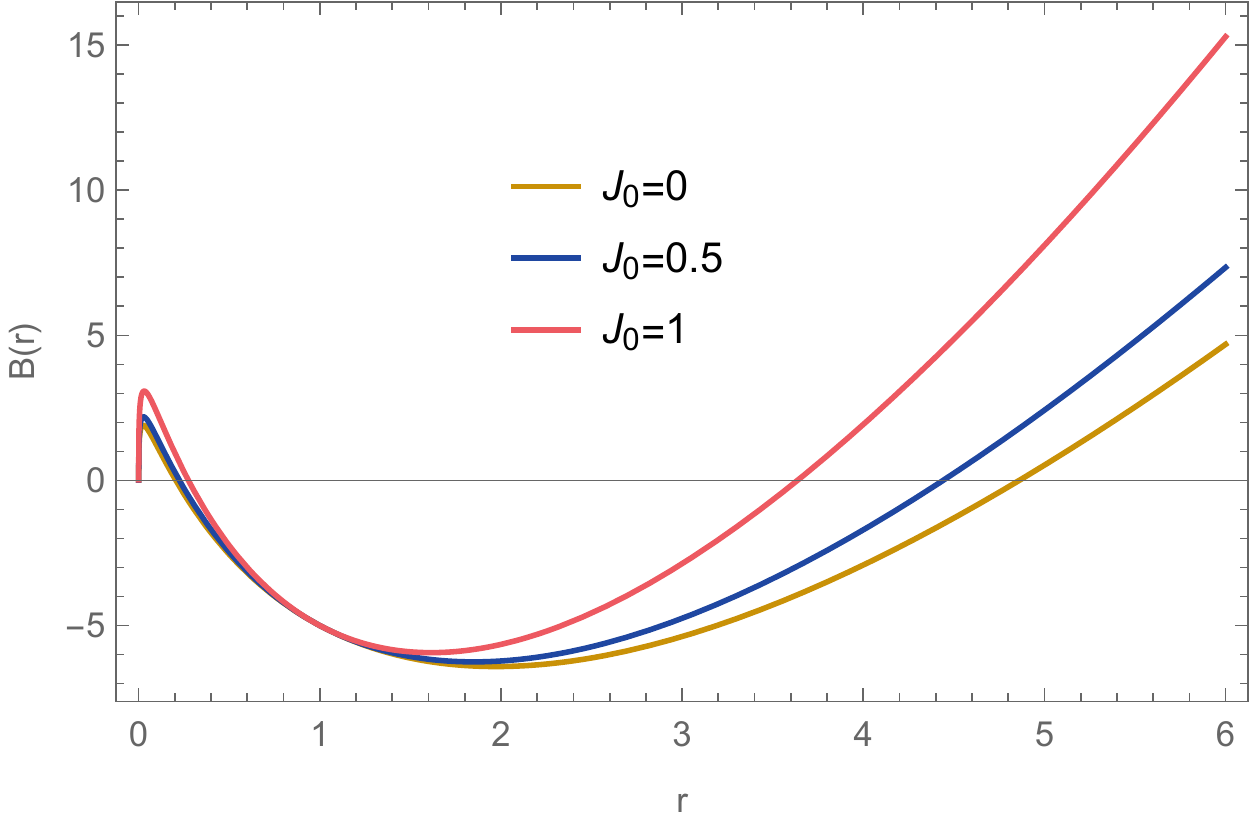}
\includegraphics[width=.40\textwidth]{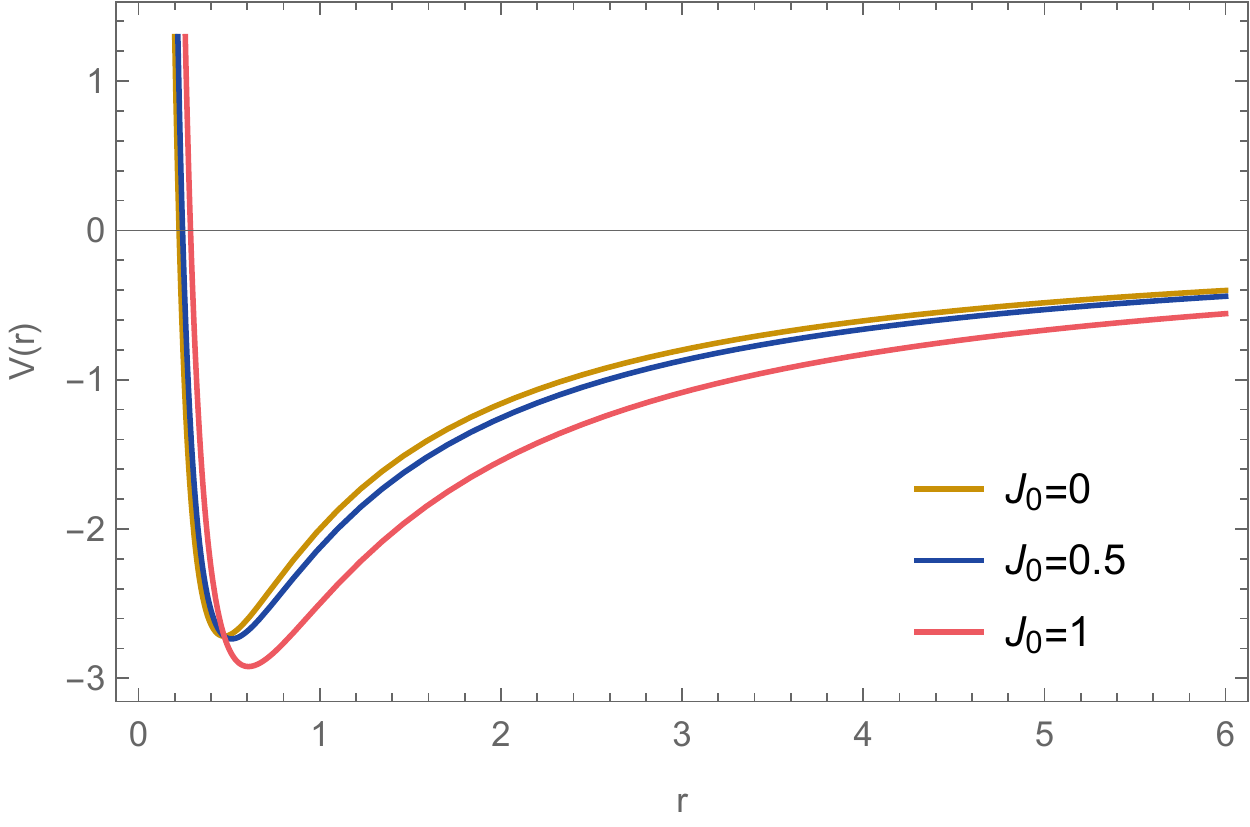}
\caption{The metric function $B(r)$ and the potential $V(r)$ versus $r$, for $\alpha=1$, having set $l=b=Q=1,M_0=5$, while changing $J_0$.} \label{a1}
\end{figure}

In the above equations, $M_0$ and $J_0$ are constants of integration.
In Fig. \ref{a1} we plot the metric function $B(r)$ and the potential $V(r)$ for different values of $J_0$, including the $J_0=0$ case in order to compare our solution with the non-rotating solution \cite{Dehghani:2017zkm}. We can see that the non-rotating black hole possesses a larger event horizon radius in comparison to the rotating case, while in the rotating case the potential forms a deeper potential well between the inner and event horizon.

We will now discuss the thermodynamics of this black hole. The Hawking temperature is found to be
\begin{equation} T(r_h) = \frac{B'(r_h)}{4\pi}=  \frac{3 (b-l Q) (b+l Q)-l^2 \left(b^2 J_0^2+2 Q^2\right) (\ln (b/r_h))}{2 \pi  b^{2/3} l^2 \sqrt[3]{r_h}}~,\end{equation}
while the extremal black hole with zero temperature has a extremal horizon at
\begin{equation} r_{\text{extremal}} = b \exp \left(\frac{3 (l Q-b) (b+l Q)}{l^2 \left(b^2 J_0^2+2 Q^2\right)}\right)~.\end{equation}
In the case with $b=Ql$, the radius will be $r_{\text{extremal}}=b$, purely determined by the scalar length scale $b$, while the entropy and the conserved angular momentum will become
\begin{equation} S(r_h) =  4\pi^2 b^{2/3} r_h^{1/3}~,\end{equation}
\begin{equation}
    J=\pi b^2 J_0~.
\end{equation}
Using the same method as before, we find that the conserved mass of the black hole cannot be finite. In particular, we can have a finite energy density, however there is a contribution in the conserved mass being related to the angular velocity
\begin{equation} 2\pi R_{j\theta}\xi^{\theta} \to 3\pi b^2J_0^2\ln \frac{R}{b}~,\end{equation}
which is divergent for the observer at large distances ($R\to\infty$).
This strange behavior is caused by the logarithmic term of the angular velocity, which is also divergent at large distances indicating that the black hole rotates faster for larger $r$. Nevertheless, this pathology can be cured by a more general solution with a scalar field $\phi(r)=\frac{1}{3} \ln \left(\frac{b}{r}+1\right)$ regular at spatial infinity and the corresponding angular velocity $u(r)=u_0+u_1 \left(\ln \left(\frac{b}{r}+1\right)-\frac{b}{b+r}\right)$ also becomes finite at large distances. In this case, the metric function and the potential can be solved explicitly but very complicated, therefore we do not show them here. Similar behaviors are also present in higher dimensional slowly rotating black holes \cite{Sheykhi:2007bv} where the solution is ill defined for $\alpha=1$. Also in \cite{Dehghani:2004sa} it is found that for $\alpha=1$, the resultant spacetime is asymptotically flat in the absence of potential, while for the particular Liouville potential the authors used, the solution breaks down for $\alpha=\sqrt{3}$ which corresponds to Kaluza-Klein theory.

\section{Conclusions}
\label{conc}

In this work we considered Einstein-Maxwell-Dilaton theory in $(2+1)$-dimensions in which the dilatonic matter is coupled to the electromagnetic field and obtained exact rotating BTZ-like solutions. There are two important parameters that they control the properties of the found black hole solutions.
The dilatonic parameter $\alpha$ which appears in the dilatonic function  coupled to the electromagnetic field, has an important impact on the metric function, affecting also the rotating properties of the black hole and also its thermodynamics. As expected, as $\alpha$ is increasing the horizon of the black black hole is increasing and also its mass. We found that the dilatonic parameter $\alpha$  does not affect the conserved electric charge, while  with the increase of $\alpha$  the angular momentum $J$ becomes larger.

The other crucial parameter that influences the properties of the black hole solution is the scalar length scale $b$ of the dilatonic matter $\phi$. Since the scalar length scale $b$ is of the secondary type, it enters in both conserved mass and the conserved angular momentum. With  increasing $\alpha$  the angular momentum can be enlarged for $b>1$ or it can be shrunk for $b<1$ indicating that
the  dilatonic matter makes the black hole rotate slower. Calculating the  entropy we found that it is always positive and the dilatonic black holes may have higher entropy than the BTZ black hole, depending on the value of the scalar length scale $b$. For $b<r_h$, where $r_h$ is the black hole horizon, the dilatonic BTZ-like black hole is thermodynamically preferred, elsewise for $b>r_h$ the BTZ black hole is preferred.

We had also discussed the particular case that arises as low energy limit of string theory, the case of $\alpha=1$. In this case the solution of the field equations is ill defined because  the angular velocity is divergent at large distances and this results in the  non-finiteness of the conserved mass. However, the black hole horizon still exists and this led the authors in \cite{Martinez:1999qi}, who studied the BTZ black hole and found a logarithmic divergence of the electric potential, to compute the mass of the black hole in a finite radius $r_0$ that encloses the black hole. In our rotating case, the  low energy limit of string theory cannot be realised unless we relax some of the assumptions we considered in order to derive the field equations.
For example,  we can introduce another degree of freedom in the metric function so that $g_{tt}g_{rr}\neq1$, or to consider another form for the potential $V$ and the circumference function $R$. We can also introduce the three form $H_{\mu\nu\rho}$ term in the action which arises from the string theory. We leave these possibilities for a future work.

\end{document}